\documentclass[fleqn,10pt]{article}
\usepackage{amsmath,amssymb}
\usepackage{graphicx}
\usepackage{epstopdf}
\usepackage[utf8]{inputenc}
\usepackage{mathrsfs}  
\usepackage{geometry}

\title{Holographic Traction Force Microscopy}

\author{Stanislaw Makarchuk$^1$, Nicolas Beyer$^1$, Christian Gaiddon$^2$,\\ Wilfried Grange$^{1,3, *}$, Pascal Hébraud${^1, *}$\\
\footnotesize{
$^1$ Université de Strasbourg, IPCMS/CNRS, UMR 7504, 23 rue du Loess,Strasbourg, 67034, FRANCE}\\
\footnotesize{$^2$ Université de Strasbourg, Inserm U1113, $3$ avenue Molière,  Strasbourg, 67200, FRANCE}\\
\footnotesize{$^3$ Université Paris Diderot, Sorbonne Paris Cité, Paris, FRANCE}\\
\footnotesize{$^*$ Corresponding authors: wilfried.grange@ipcms.unistra.fr, pascal.hebraud@ipcms.unistra.fr}}

\date{}
\begin{document}
\maketitle

\begin{abstract}
Traction Force Microscopy (TFM) computes the forces exerted at the surface of an elastic material by measuring induced deformations in volume. It is used to determine the pattern of the adhesion forces exerted by cells or by cellular assemblies grown onto a soft deformable substrate. Typically, colloidal particles are dispersed in the substrate and their displacement is monitored by fluorescent microscopy. As with any other fluorescent techniques, the accuracy in measuring a particule's position is ultimately limited by the number of evaluated fluorescent photons. Here, we present a TFM technique based on the detection of probe particle displacements by holographic tracking microscopy. We show that nanometer scale resolutions of the particle displacements can be obtained and determine the maximum volume fraction of markers in the substrate. We demonstrate the feasibility of the technique experimentally and measure the three-dimensional force fields exerted by colorectal cancer cells cultivated onto a polyacrylamide gel substrate. 
\end{abstract}

\section*{Introduction}

Cells exert forces between each other and onto their environment. When cultivated \textit{in vitro}, cells exert forces onto the culture substrate. These forces are generated by the actin-myosin network, in association with proteins to induce adhesion onto the cell environment.  Among them, integrins are responsible for cell/extracellular matrix adhesion, and, cadherins  for cell/cell junctions. Cellular forces are not spatially homogeneous; for instance, when cells are cultivated onto a flat substrate, forces mainly occur at localized regions, called focal adhesion sites. These regions involve several tens of proteins~\cite{Wolfenson2009}. Both focal adhesion sites sizes and shapes strongly depend on the physiological context. The adhesion stress pattern between neighboring cells is different from that involving the interaction between a cell and the extracellular matrix~\cite{Cukierman2001a}. 
It has been also observed that mechanical properties play a key regulation role in many cellular processes~\cite{Discher2005a}, not limited to migration. The link between the mechanical phenotype of cells and the onset of diseases (e.g. cancer) is a subject of a considerable interest~\cite{agus_physical_2013};  a change in mobility allows a single cell to detach from a primary tumor site, infiltrate adjacent tissues, penetrate the vascular walls and finally colonize competent organs.

To understand the roles of specific molecular processes in the mechanical phenotype of cells, it becomes necessary
to measure precisely how the expression of specific proteins changes the forces exerted by the cells on their
environement. Several techniques have been developed to measure the adhesion forces generated by cells onto their environment~:  micropipette aspiration ~\cite{hochmuth_micropipette_2000} and flow techniques ~\cite{lu_microfluidic_2004} measure the overall value of the forces exerted by a cell in response to an external stimulus. 
Similarly, several methods have been developed to study the forces exerted by a cell on a soft substrate. They can be classified as follows: {(i)} The measurement of the deformation of an elastic substrate. These studies, pioneered by Harris~\cite{Harris1980,Harris1981}, consist in analyzing the wrinkling pattern induced by the application of forces onto a thin elastic silicone sheet. Because there are no simple ways to convert wrinkle patterns into a traction forces map, this method remains qualitative and is not used nowadays. {(ii)} The force measurements based on  growing cells onto an array of pillars, acting as force sensors. The measurement of the deformation of each pillar allows the determination of the applied force~\cite{Galbraith1997, tan_cells_2003,du_roure_force_2005, saez_rigidity-driven_2007}. The force can be easily calculated using Hooke's law for each pillar. While being widely used, this method has significant drawbacks : the non-physiological shape of substrates might affect cellular responses. Moreover, cellular shapes are strongly affected by both the dimensions of the micro-pillars and the mesh size ~\cite{tee_cell_2011}. {(iii)} The measurement of the three-dimensional field of deformation of a soft substrate embedded with particles~\cite{Dembo1999, del_alamo_three-dimensional_2013} - Traction Force Microscopy (TFM). From the displacement of the probe particles, forces can be determined.\\

In TFM, the surface force field $\mathbf{F}(\mathbf{r'})$ at the surface of the substrate is computed from the elasticity equation, $\mathbf{u} = G \mathbf{F}$, where  $\mathbf{u}(\mathbf{r})$ is the displacement field and $G$ the Green function.  As a consequence, an inverse problem has to be solved: the forces at points at the substrate surface, $\mathbf{r'}$, must be computed from the knowledge of the displacement field at a given set of points $\mathbf{r}$ inside the elastic substrate. 

A direct solution of the elasticity equation could be obtained using Singular Value Decomposition of the matrix $G$ but the condition number of $G$ is very high (typically $10^3$). This implies that the addition of force values onto the direction defined by the lowest singular values of $G$ would induce negligible change in the overall displacement $u$. Therefore,  the addition of a small noise to the measured displacement field significantly alters the computed values of the force field;~the problem is ill-posed. Several strategies, requiring prior information, have been proposed to solve ill-posed problems (either in real or Fourier spaces~\cite{Butler2002}). For instance, regularization techniques consist in selecting a solution among the many possible and indistinguishable solutions of the ill-posed problem by imposing a penalty to solutions that exhibit some property.

When calculating the force, one can either assume that the force is highly localized (and so the force is calculated at specific points for Traction Reconstruction with Point Forces (TRPF)~\cite{Schwarz2002, Sabass2008} or that the force is distributed on a specific area (focal adhesion). 
In the latter approach, the density of markers has to be kept high enough to prevent aliasing (which would result in an underestimation of the force). In contrast, TRPF has to be performed at low particles densities. TRPF successfully recovers forces if only particles at a sufficient distance from adhesion points are considered so that dipolar and higher-order terms can be neglected. Obviously, obtained forces only represent an average on the focal adhesion but this averaged quantity (as it would be determined in experiments where microfabricated pillars support cells~\cite{du_roure_force_2005}) is sufficient to evidence for different mechanical phenotypes. Let us also mention that, for both approaches, the accuracy of particle position might also contribute in the determination of traction peaks. However, current TFM setups operate at high densities (at a few particles per  $\mu m^2$,  see below) so that noise field in the displacement field only contributes very little the quality of the reconstruction~\cite{zundel_factors_2017}.

Nowadays, state-of-the art TFM instruments are fluorescence based-devices that aim at determining the fine structure of small focal adhesion. These devices use either discs composed of quantum dots ~\cite{bergert_confocal_2016} or beads having diameters of few tens of nm~ \cite{colin-york_super-resolved_2016}. Electrohydrodynamic nanodrip-printing of quantum dots allow placement of the discs at very specific positions and so well defined patterns can be realized  (grid size of $1.5$ micrometers, printing error of $35$ nm). Higher densities can be achieved with beads ($2.2$ $\mu m^{-2}$ for beads of $40$ nm in diameter) using Stimulated Emission Depletion Microscopy (STED) that confines the fluorescence emission to a region much smaller that the typical (diffraction limited-) fluorescence spot (Super-Resolved Traction Force Microscopy (STFM)~\cite{colin-york_super-resolved_2016}). While STFM dramatically improves the sensitivity of TFM (albeit currently limited to 2D measurements) and so offers an attractive alternative to current (low resolution) TFM fluorescence based devices ~\cite{colin-york2018}, it suffers from severe limitations. STFM equires expensive and sophisticated optical setups (and this explains why STED is certainly less established than Photo-activated localization microscopy (PALM) or Stochastic Optical Reconstruction Microscopy (STORM)~\cite{huang_three-dimensional_2008} in Biology), is limited to the imaging of thin gels (to reduce optical aberrations~ \cite{colin-york_super-resolved_2016}) and, more importantly, can induce photodamage (as high-power depletion lasers are used for periods (hours for TFM) that far exceeds those used in conventional STED experiments)\cite{waldchen2015}.

In studies where it is sufficient to determine forces without the need to resolve focal adhesion (e.g. identifying metastatic and non-metastatic cell lines based on their capabilities to exert large traction forces~\cite{wirtz2011}), low particle densities could be used. There, however, it remains crucial to resolve displacements with nm-accuracy in all three directions. As the number of collected photons is usually low in fluorescence measurements (\textit{i.e.} shot noise is severely limiting), the typical accuracy of conventional fluorescence-based TFM devices (about $4$ to $8$ nm for in-plane and $20$ nm for out-of-plane measurements when using Qdots, respectively ~\cite{bergert_confocal_2016}) certainly would fail to recover forces in all three dimensions. 

In this paper, we present a novel approach, which consists in monitoring the displacements of non-fluorescent micrometer-sized particles at low spatial density. Using a slightly coherent light source (Light Emitting Diode, LED), we analyze the diffraction patterns that originate from
the interference between the scattered and the incident rays and reach nm localization accuracies along all three directions ~\cite{gosse_magnetic_2002}. To highlight this new method, we present non-filtered traction forces maps obtained for a colon carcinoma cell line (SW480) seeded on a polyacrylamide gel. In particular, we report on out-of-plane forces that are rarely measured in fluorescence-based TFM but are known, however, to play a significant role in both cell adhesion and migration ~\cite{delanoe-ayari_4d_2010}.

\section*{Displacements of the particles and noise analysis}

\subsection*{Positions of the particles}

\begin{figure}[h!]
      \begin{center}
              \includegraphics[width=0.7\textwidth]{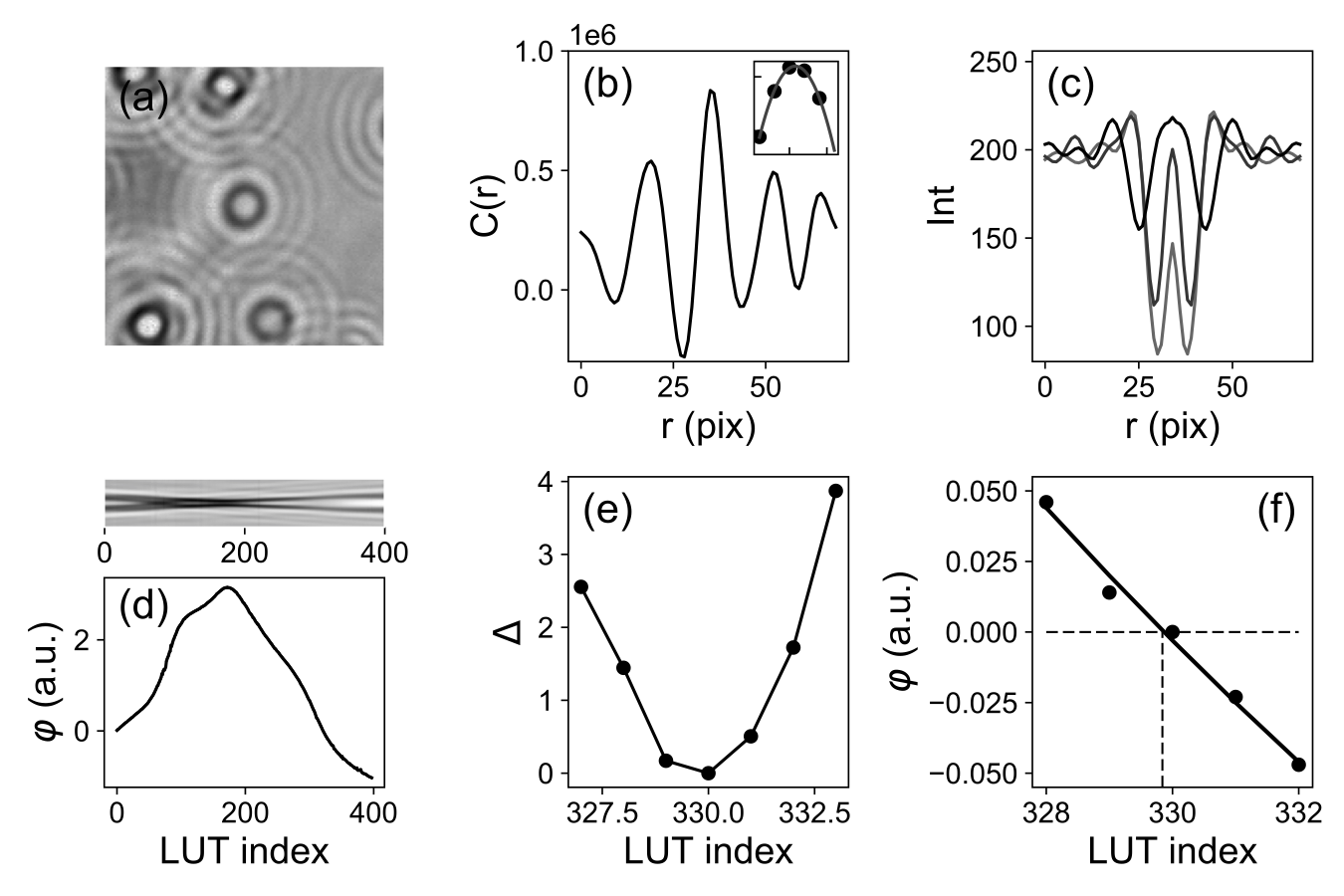}
      \end{center}
\caption{Tracking the positions of particles along $x$, $y$ and $z$. \textbf{(a)} Typical image of beads embedded in a polyacrylamide gel and imaged with a LED (141x141 pixels, 110 nm per pixel). \textbf{(b)} Knowing the approximate position of the bead (here  at the center of the image), a 1D cross correlation ($C(r)$) can be calculated for both $x$ and $y$. Then, a  polynomial fit (inset) around the maximum of $C(r)$ allows to determine the position with sub-pixel accuracy  (see Supplementary Information). \textbf{(c)} $2$-sided intensity profiles obtained at different positions along the $z$ axis (position: $6.5\ \mu$m, $11.5\ \mu$m and $16.5\ \mu$m relative our $z$ reference). \textbf{(d)} From these profiles, a Look Up Table (LUT) is built (shown here for 400 $z$-positions, step size $50$ nm) and allows to calculate the parameter $\varphi$ that varies quadratically with the $z$ position. \textbf{(e)} To determine the $z$ position of a particle (its index in the LUT), squared differences ($\Delta$) are calculated (between the profiles form the LUT and that of the particle of interest). \textbf{(f)} A quadratic adjustment of the phase \textit{vs} the LUT index around the minimum of $\Delta$ allows to determine the $z$ position with a high precision (the index at which the phase vanishes; here $329.84$ corresponding to a $z$ position of $16.4925\ \mu$m). See~\cite{gosse_magnetic_2002} and~\cite{cavatore_microscopie_2011} for more details.}
\label{fig:tracking}
\end{figure}

We use robust algorithms to determine the $x$ and $y$ (in-plane) and $z$ (out-of-plane) positions of each particle (Fig.~\ref{fig:tracking} \textbf{(a)}). At high Signal to Noise Ratio (SNR) and high magnification ($50$ and above), these algorithms, which are mainly used in single-molecule experiments (magnetic tweezers \cite{lionnet_single-molecule_2012}), are capable of determining the position with a precision better than  $1/100$th of a pixel (in $x$ and $y$) and below 1 nm in $z$. To this end, we first compute a 1D cross correlation (Fig.~\ref{fig:tracking} \textbf{(b)}) to determine the centers of the particles. Then, we compute an intensity profile (the average intensity of pixels located at a given distance from the center)~\cite{zhang_three-dimensional_2008}, which is subsequently compared with a calibration table. This Look Up Table (LUT) is obtained from intensity profiles measured at known and given distances, e.g.  by moving the  objective lens every 50 nm and averaging a series of images,  (Fig.~\ref{fig:tracking} \textbf{(d)} top). To obtain an accuracy better than the objective step size, we calculate the  squared differences (between the radial profile intensities of the measured particle and those of the LUT) and perform a Least Squares polynomial  adjustment (Fig.~\ref{fig:tracking} \textbf{(e)}). Discretization errors have to be taken into account (similar errors occur for $x$ and $y$ when sampling is poor, \textit{i.e.} when the pixel size is not small enough with respect to the diffraction pattern features \cite{vanloenhout_non-bias-limited_2012}). To correct for a  possible bias,  we follow the ideas of Gosse and Croquette~\cite{gosse_magnetic_2002} and estimate a parameter (a phase computed from the Hilbert transform of the intensity profile), which has a known (quadratic) dependence with the $z$ position (Fig.~\ref{fig:tracking} \textbf{(f)}) (see also~\cite{cavatore_microscopie_2011}). Other approaches  have been proposed to reduce discretization errors: Cnossen et al.~\cite{cnossen_optimized_2014} use an an iterative approach and have assumed a linear dependence of the bias with the position. We have found, however, that this approximation was somehow arbitrary and fails to correct for the bias (see Supplementary Information). Finally, and in agreement with previous studies~\cite{cavatore_microscopie_2011}, we have found that these algorithms are limited by shot noise (so tracking noise scales as 1/ $\sqrt{N}$, where $N$ is the number of evaluated photons) and that photon noise dominates quantization noise so that it is sufficient to work with 8 bit-images.

\subsection*{Localization accuracy}

As the noise level varies with bandwidth, the Allan deviation (AD, which measures the noise level $\sigma$ when averaging over a given bandwidth $1/\tau$) is a relevant parameter to estimate the noise level of the instrument~\cite{allan_statistics_1966, czerwinski_quantifying_2009}. The AD allows to distinguish between different types of noise. For instance, tracking noise decreases by a factor $\sqrt{n}$ when averaging over $n$ measurements. In contrast, thermal drift will cause a monotonic increase in the AD at large enough $\tau$ ($\sim$ 0.5 s and above) and environmental noise shows an increase in the AD over a specific range of times only (below a fraction of a second). Assuming only two sources of noise in a given time measurement (tracking noise and thermal drift), we then expect the AD to decrease at low $\tau$ (0.5 s and below) and then to increase at higher $\tau$. Note that the AD is expected to be noisy when  $\tau$ approaches the total measurement time (simply because less data are averaged). Obviously, with the addition of correlated noise (e.g. fans), we may observe different patterns at intermediate $\tau$ that simply correlate with the magnitude of the different noise levels. 

As shown in Fig~\ref{fig:Allan} \textbf{(a)}, the AD (measured on a $1$ micrometer particle embedded in the gel)  is much larger on $z$ (as compared to $x$ or $y$). This behavior is somewhat expected as we use a LED with a low coherence ($\sim$ 6 $\mu m$ only, to prevent overlapping of diffraction patterns from adjacent particles; see below). For this light source, the radial profiles (see above) are less spatially extended and poorly defined~\cite{dulin_efficient_2014}. This results in a low performance of the algorithm along $z$ and so tracking noise possibly dominates over environmental noise  (in the range 0.05 to 0.5 s). In contrast, cross-correlation gives satisfying results at low SNR~\cite{vanloenhout_non-bias-limited_2012} and an opposite behavior is observed in $x$ and $y$.
At high $\tau$ (in the range 0.8 to 2s), the AD shows an increase in all directions that can then be correlated with thermal drift. However, that drift (and correlated noise) can be significantly reduced with fiduciary markers, by subtracting the position of reference particles fixed on the glass surface from those measured (Fig~\ref{fig:Allan} \textbf{(b)}). As expected, random (tracking) noise can be reduced by averaging only and this explains why a doubling of the AD is observed at low $\tau$ when calculating position differences ($z$ direction). When calculating such differences, we observe that averaging positions of single images gives similar results as averaging images and then calculating a position ( Fig~\ref{fig:Allan} \textbf{(b)}, disks). For computation requirements, we have chosen the latter approach. 

Note finally that it exists region, from $2$ to $6$ micrometers below the focus, where the tracking on $z$ is more accurate (Fig~\ref{fig:Allan} \textbf{(b)}, \textit{inset}). This is in agreement with previous results~\cite{cnossen_optimized_2014} and should be attributed to the fact that the slope of the 
phase difference between adjacent LUT planes (see above) depends on the distance from the focus. Again, the cross correlation is more robust and the $x$ and $y$ positions show almost no dependence on the particle position along $z$. 

Altogether, the results of this analysis show that we can measure the displacements of probe particles with an accuracy of $\sigma_{xy} \sim 1$ nm and  $\sigma_{z} \sim 3$ nm, an order of magnitude better than what has been reported in recent, state-of-the art fluorescence-based TFM measurements.

\subsection*{Spatial resolution}

The measure of the displacement field is performed at randomly positioned particles.  To increase the resolution of the displacement field, one has to increase the number of particles \textit{per} unit volume that can be tracked. This concentration is limited by the volume of the diffraction pattern of each particle, which depends on the size and optical index of the particles as well as on the wavelength and the spectral width of the light source. As it may be difficult to resolve 
the positions of the particles  whose diffraction patterns intersect, we expect the optimum particle concentration to be of the order of the ratio between the volumes of a particle and that of the diffraction pattern. Assuming  a cone ($10$ $\mu$m for the height and $6$ $\mu$m for the base diameter) and a particle diameter of $1\ \mu$m, the optimum volume fraction is expected to be $c_{max}=\frac{4/3 \pi (0.5)^3 \mu m^3}{1/3 \pi 3^2 \mu m^2 10 \mu m} \approx 0.056= 0.56 \%$. 

\begin{figure}[h!]
      \begin{center}
              \includegraphics[width=0.7\textwidth,]{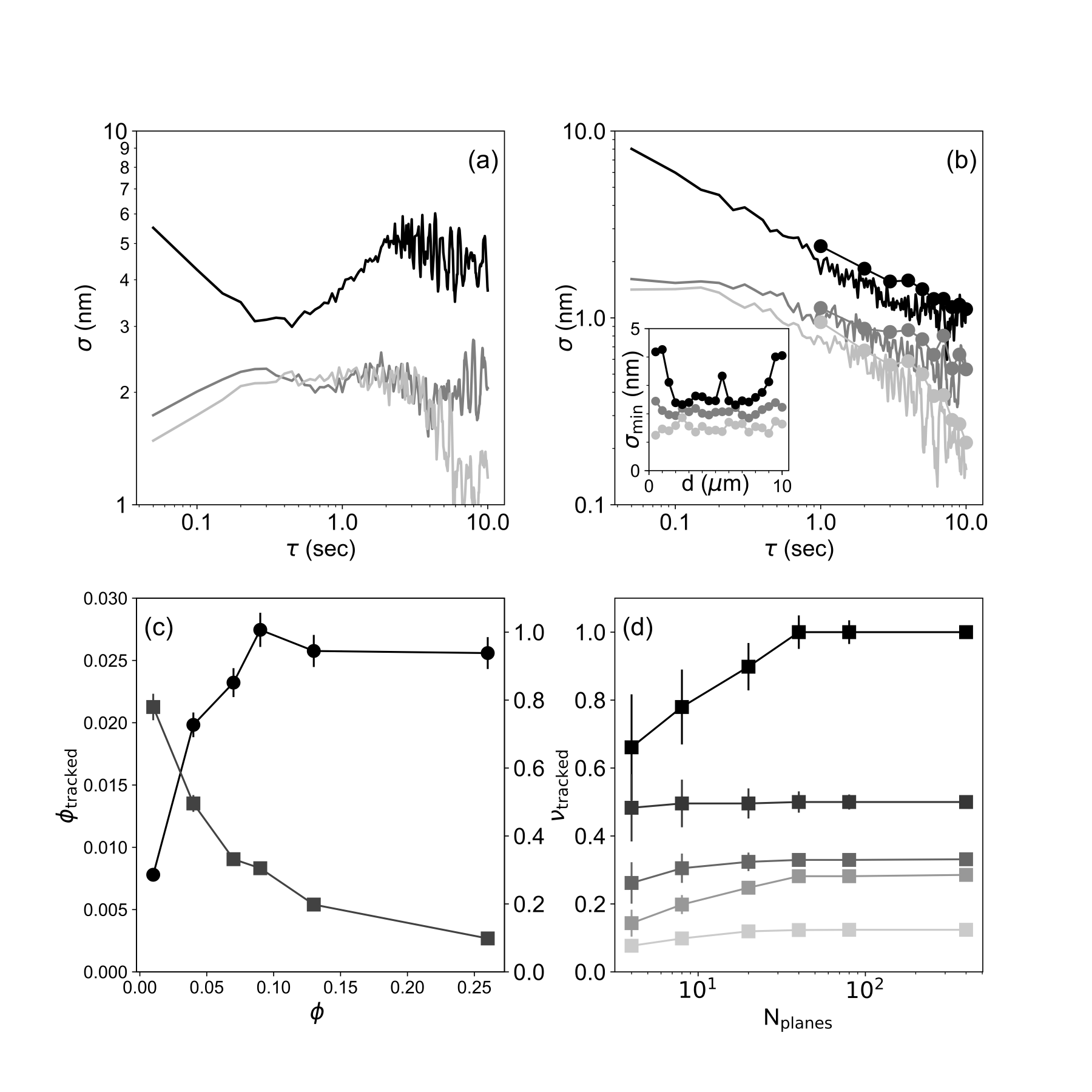}
            \end{center}
\caption{Top row : noise measurements for a one micrometer particle embedded in a polyacrylamide gel.\textbf{(a)} Allan deviations (AD, sampling frequency: $20$ Hz) measured for all three directions ($x$: light grey, $y$: dark grey and $z$: black).\textbf{(b)} When subtracting reference particles (melted on a glass surface), the AD is strongly reduced at high $\tau$ values (0.5 s and above). Note that averaging images (circular markers) or averaging positions (lines) give similar results (see main text). \textit{Inset}: AD $versus$ vertical distance to the focus. For $z$, there are optimal positions for tracking (from about $2$ to $6$ micrometers below the focus). The $x$ and $y$ positions are less sensitive to the diffraction patterns and the AD remains constant.
Bottom row : optimal particle volume fraction. \textbf{(c)} Evolution of the volume fraction of successfully tracked particles, $\phi_{tracked}$ ($\phi$ (black disks, left axis)), and of the relative fraction of successfully tracked particles, $\nu_{tracked}$, (grey square symbols, right axis) as a function of the particle volume fraction. \textbf{(d)} Relative fraction of successfully tracked particles as a function of the number of acquired planes, $N_{planes}$, for different particles volume fraction. From top to bottom~ : $\phi=0.01\ \%$, $0.04\ \%$, $0.07\ \%$, $0.13\ \%$, $0.26\ \%$. From $5$ to $8$ independent regions of size $60X60\ \mu$m$^2$ are observed at each particle volume fraction. Error bars correspond to one standard deviation of these measurements. }

\label{fig:Allan}
\end{figure}

Experimentally, we determine the optimal particle concentration as follows~: $800$ frames at different $z$ positions are acquired, using a step size $\Delta z = 25$ nm between two frames. This ensemble of images is divided into two subsets~: the first one contains images obtained at positions $2i\Delta z $; the second contains images acquired at $(2i+1)\Delta z $ ($i=0:399$). The first set is used as a LUT (with a step size $\Delta z_{LUT} = 50$ nm) for tracking particle positions from the second set of images. As the true positions of the tracked images (corresponding to $2(i+1)\Delta z $) are known, we can determine whether they are correctly tracked relative to the other subset of images. This procedure mimics what is obtained in a real experiment as the tracked beads are more likely to be found in between two LUT planes. We then select $8$ planes  from the tracked planes subset, corresponding to a reasonable number of planes acquired in a real experiment. We compute the $x$, $y$ and $z$ positions of the particles. A particle is considered to be successfully tracked when the following conditions are met: (i) the exact same $x$ and $y$ positions ( $\pm 11$ nm, which is one tenth of a pixel or about $3$ standard deviations) have to be obtained for at least two tracking planes and (ii)  the $z$ position should be found within the same LUT intervals ($\Delta z$).
Fig.~\ref{fig:Allan} shows the result of this analysis. Due to the increasing number of diffraction pattern overlaps, the relative fraction of successfully tracked particles, $\nu_{tracked}$, defined as the ratio between the number of tracked particles over their total number, decreases with the volume fraction (Fig.~ \ref{fig:Allan} \textbf{(c)}, squares). The volume fraction of successfully tracked particles, $\phi_{tracked}$, increases with particle concentration and saturates when the diffraction patterns from different particles overlap (Fig.~ \ref{fig:Allan} \textbf{(c)}, circles). This results in a maximum volume fraction of successfully tracked particles, $\phi_{max}=0.1\ \%$, that we define as the optimal particle volume fraction.\\
Obviously, the fraction of successfully tracked particles depends on the number of planes that are imaged ( Fig.~\ref{fig:Allan} \textbf{(d)}). When the number of tracking planes decreases, more particles are not detected as the probability to track a particle in the optimal tracking region $d_z$ (from $2$ to $6\ \mu$m below the focus,  Fig~\ref{fig:Allan} \textbf{(b)}, \textit{inset}) decreases.

The above analysis sets the maximum spatial resolution of our apparatus~:  the maximum volume fraction of successfully tracked particles is $\phi_{tracked}= 0.037\ \%$, for an optimum concentration of particles $\phi=0.13\ \%$, and $40$ tracking planes. This corresponds to an average distance between the centers of the particles of $ 6.7\ \mu$m. Nevertheless, and to reduce computation time, we will image only four planes and use a particle concentration  $\phi$ of $0.1$. Under these conditions, the volume fraction of successfully tracked particles is $0.027\ \%$ and the average interparticle distance is $7.5\ \mu$m.  Note that an appropriate patterning of beads,  preventing overlapping of in-plane and out-of-plane diffraction fringes, would allow a larger fraction of beads to be tracked~\cite{vlaminck2011}. Because our algorithm does not require a high density of beads, the fact that a large fraction of beads cannot be tracked (roughly a factor $4$ when knowing the true position of beads and setting a cut-off at one standard deviation from the true position) does not represent a significant obstacle to the successful  reconstruction of traction forces. 

\begin{figure}[h!]
      \begin{center}
              \includegraphics[width=0.7\textwidth]{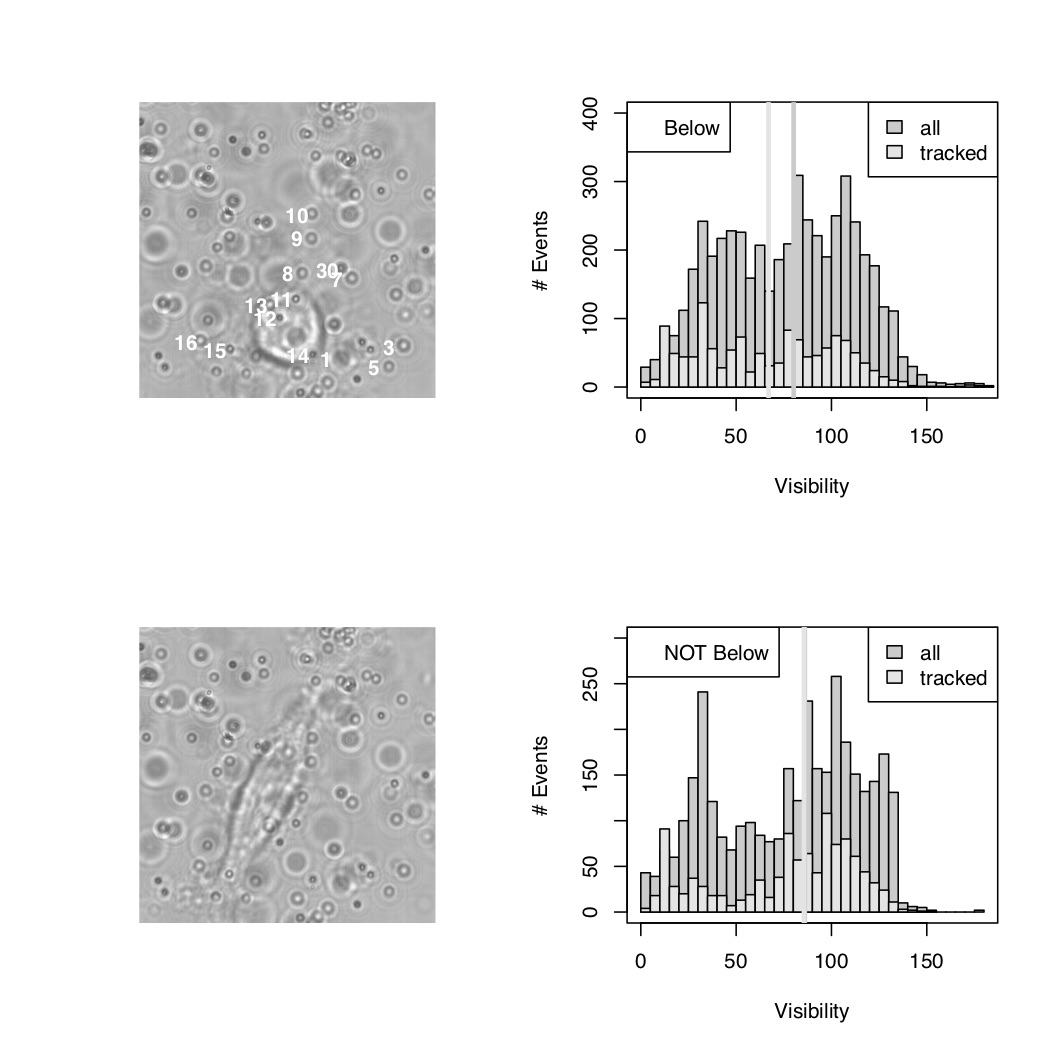}
            \end{center}
\caption{Left: Typical images (550 x 550 pixels) of a cell seeded onto a polyacrylamide gel. Images have been obtained  130 (top image) and 260 (bottom image) minutes after seeding. Shown also (top image, white) are the identification numbers of 14 particles, which have been found below the cell in the time course (600 minutes) of the experiment. Right. Histograms of  the visibility (defined in the $z$ intensity profile as the difference between the maximum of the first peak and the minimum of the first valley) when beads are located below (top) and  not below (bottom) the cell. At each location, the histograms have been computed when the beads are successfully tracked (light grey) and are either successfully or not successfully tracked (grey). The vertical lines indicate the value of the second quartile of the distributions. }
\label{fig:visibility}
\end{figure}

Two key properties of tracking techniques in TFM are their ability to track beads underneath cells and their accuracy. Fluorescence techniques suffer from the sensitivity of cells to the relatively large intensities necessary to excite the tracker bead's fluorescence and from the intrinsic cell fluorescence. Here, the diffraction patterns are altered by refraction of light by the cellular organelles. 
To determine whether tracking is influenced by the presence of the adhered cell, we compute the visibility (defined as the difference in the radial profile between the maximum of the first peak and the minimum of the first valley~\cite{dulin_efficient_2014}) for a series of beads (14) that may be below or out of the cell during an experiment (Fig.~\ref{fig:visibility}). As several planes can contribute to tracking (\textit{i.e.} resulting in an identical index in the LUT), we have chosen to use the maximum of the obtained visibilities when a particle is successfully tracked. Comparing the obtained distributions when the particles are either below or not below the cell allows us to determine whether the accuracy (which correlates with the visibility~\cite{dulin_efficient_2014}) is modified by the presence of the cell. As shown in Fig.~\ref{fig:visibility}, the obtained distributions (light grey) result in different medians (vertical lines, light grey). We find values of  ($77\ \pm$ 2) and   ($66\ \pm$ 2) ($0.95$ confidence interval) for the estimates of the means and a Wilcoxon Rank-Sum test indicates that the distributions are statistically different (p-value less than $1.2\cdot 10^{-12}$). In addition, the probability to successfully tracking a bead depends on its location. We find values of ($0.32\ \pm 0.03$ and $0.26\ \pm 0.07$) ($0.95$ confidence interval) for the estimates of the probabilities of tracking a particle when not below and below a cell. Again, these values are  are statistically different (p-value of  $2.6\cdot 10^{-7}$, chi-squared test). When comparing the overall visibilities (the maximum of the visibilities determined at the four different tracking planes, independently of the success of the tracking) with the previously obtained values (when tracking is successful), we found no statistical difference when particles are not below the cell (p-value larger than $0.14$, Wilcoxon Rank-Sum test) but a statistical difference when particles are below the cell (p-value less than $2.3\cdot 10^{-16}$, Wilcoxon Rank-Sum test). This finding also indicates that the cell presence of the cell indeed lowers the visibility value, but that when a cell passes our criteria for tracking, the visibility is similar below and not below the cell, indicating that the accuracy of the tracker beads mouvements are the same, whatever the relative position of the beads and  the cell .

\section*{Computation of the force field}
The deformation of the gel substrate is assumed to be small enough so that linear elasticity theory can be used. The traction force field is calculated from the measured displacement field by inverting the elasticity equation. We need to solve  $\mathbf{u}=G \mathbf{F}$ where $\mathbf{u}=(u_x(\mathbf{r}^{(1)}), u_y(\mathbf{r}^{(1)}), u_z(\mathbf{r}^{(1)}), u_x(\mathbf{r}^{(2)}), u_y(\mathbf{r}^{(2)}),...) $ and   $\mathbf{F}=(F_x(\mathbf{r}'^{(1)}), F_y(\mathbf{r}'^{(1)}), F_z(\mathbf{r}'^{(1)}), F_x(\mathbf{r}'^{(2)}), F_y(\mathbf{r}'^{(2)}),...)$  are $1$D-vectors of the displacement field, measured at positions $\mathbf{r}$ inside the gel and of the force field, measured at positions $\mathbf{r'}$ at the gel surface.  $G$ is a $2D$-matrix.  For $N$ displacements and  $M$ force points, the size of the displacement vector (at 3 dimensions) is $3N$, the size of the force vector is $3M$ and the Green matrix  has a size of $3N$x$3M$, respectively.  

When the substrate is thick enough (larger than the characteristic depth of the deformation induced by cell adhesion), the elastic medium can be considered as semi-infinite and the Green function is that of a semi - infinite medium~\cite{LANDAU1976}, given by Boussinesq. The Poisson ratio of polyacrylamide gels being close to $0.5$, the elements of the matrix $g$ are given by~: 

\begin{equation}
g_{kl}(R)= \frac{3}{4 \pi E R^{3}}(\delta_{kl}R^{2}+R_k R_l)
\label{green}
\end{equation}

\noindent where $E$ is the Young's  modulus of the medium and $R=| \mathbf{r} - \mathbf{r'}|$ the vector between the displacement point and the force point. Note that taking a value of 0.5 for the Poisson ratio is an approximation, which is commonly made in TFM measurements. However, an error in its determination could affect the estimated forces. A possibility would consist in the direct determination of the ratio using new technqiues like  two-layer elastographic TFM experiments~\cite{alvarez-gonzalez_two-layer_2017}.

The Green matrix of the entire system, which relates the whole displacement field with the force field, is constructed by blocks consisting of matrices $g$ for all possible pairs of points:

\begin{equation}
G_{ij}(\mathbf{r}^{(1)},...,\mathbf{r}^{(N)}, \mathbf{r}'^{(1)}, ..., \mathbf{r}'^{(M)})=g(|\mathbf{r}^{(i)}-\mathbf{r}'^{(j)}|) 
\label{GG}
\end{equation}

The conditional number of $G$ is larger than $10^3$ and the inversion of the elasticity equation is an ill-posed problem. Therefore, some regularization is required. As stated previously, regularization consists in adding some constraints that filter out solutions that do not fulfill \textit{a priori} conditions~\cite{kirsch2011introduction}. Here, we use Tikhonov regularization \cite{tikhonov1977solutions} for which the constraint consists in introducing an expected solution $F_0$. The sum of two norms is minimized: the residual  and the divergence between the calculated and the expected solutions. It is given by:

\begin{equation}
\mathbf{F}_{reg}=min_ {\mathbf{F}} \Big(  |G \mathbf{F} - \mathbf{ u}|^2 + \lambda^2 |\mathbf{ F} - \mathbf{ F}_0|^2 \Big) 
\label{eq:tikh}
\end{equation}

Here $\lambda$ is the \textit{regularization parameter}, which weighs the regularization term $ |\mathbf{ F} - \mathbf{ F}_0|^2$. We use the \textit{L-curve} criterion \cite{hansen_l-curve_1999}, which is a log-log plot of residual norm $ |G \mathbf{ F} - \mathbf{ u}|^2 $ as a function of $ |\mathbf{F} - \mathbf{F}_0|$ for different $\lambda$. This plot exhibits an L - shape, and its corner determines the balance between data agreement and regularization. The value of the regularization parameter $\lambda$  that corresponds to this corner is chosen for  the regularization procedure. 

In the case of TFM, it is difficult to predict some force field $\mathbf{F_0}$ and the main constraint consists in setting that the traction forces should not be unreasonably large \cite{Schwarz2015}. We thus perform a zero-order Tikhonov regularization $\mathbf{F_0} =\mathbf{ 0}$. In this case, equation~\ref{eq:tikh} rewrites as: $ \mathbf{ F}_{reg}=min_ {\mathbf{ F}} \Big(  |G \mathbf{ F} - \mathbf{ u}|^2 + \lambda^2 |\mathbf{ F} |^2 \Big) $. The regularization is performed with a MatLab routine written by P.C. Hansen~\cite{hansen_2007_regularization}.

\begin{figure}[h!]
      \begin{center}
             \includegraphics[width=0.7\textwidth]{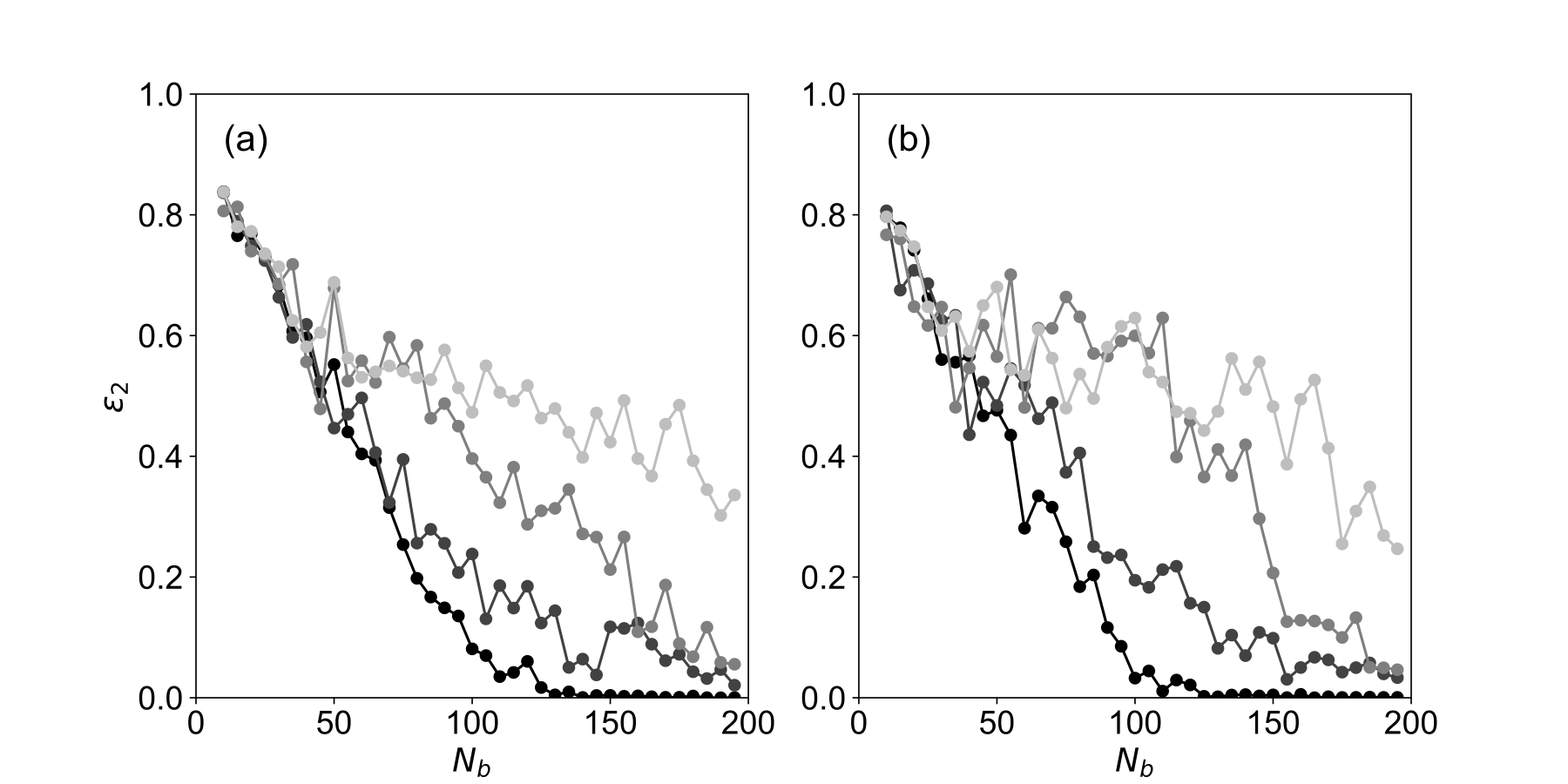}
            \end{center}
\caption{Simulation data. Error of the computed force field, $\epsilon_2=\frac{\vert F_{\mathrm{reconstructed}}-F_{\mathrm{real}}\vert}{\vert F_{\mathrm{real}}\vert}$ as a function of the number of points of measurement of the displacement, $N_b$. The number of points of reconstruction of the force field is $100$. The  beads are randomly dispersed in an elastic medium ($E=500$ Pa) of size  $100$X$100$X$50$ $\mu m^3$. A $10$ nN point force parallel  \textbf{(a)} and  normal \textbf{(b)} to the surface is applied at the center of the considered region. The displacement field is computed at each bead position. A random noise of standard deviation $\sigma$ along the $x$ and $y$ directions and $\sigma$ along the $z$ direction is added to thedisplacement field. The force field is computed at the surface of the gel ($100$ points of calculation), and the error between the computed and the applied forces is computed. Each value is an average over $10$ random distribution of particules inside the medium. From black to light grey, the added noise standard deviation is $\sigma=0$, $\sigma = 1$ nm,   $\sigma = 2$ nm and $\sigma = 10$ nm.}
\label{fig:simulforce}
\end{figure}

Although the displacement field is measured at low spatial density, the resolution of their displacement allows for the reconstruction of the force field when the number of points where the force is computed is approximately equal to the number of points of measurement of the displacement field. In Fig.~\ref{fig:simulforce}, numerical simulations of force field reconstruction are performed. A single point force is applied at the surface of the gel,  the displacement field is calculated at $N_b$ points inside the gel and a random gaussian noise is added to the computed displacements.  The force field is then reconstructed and the difference between the applied and the reconstructed force field is plotted as a function of $N_b$ for different amplitudes of the noise. One obtains that when the number of points of calculation of the force field is equal to $N_b$, the error over the reconstructed force field is lower than $20\ \%$. This accuracy is similar to that obtained when the force field is obtained in the mostly used regime where the beads displacement accuracy is lower, but the density of markers is higher, as achieved with fluorescent particles~\cite{Sabass2008}.

\section*{Results}

Adhesion experiments are performed using colorectal cancer cell line \textit{SW 480} grown onto the polyacrylamide gel (see section~\ref{sec:matmeth}). The positions of the particles are tracked for 10 hours. The time-step between two measurements is $\delta t = 60$ sec. Radial profiles of the particles at reference positions are acquired after the cells are injected inside the chamber and before they start to adhere to the substrate. We thus have access to the positions of the tracking particles in the absence of applied forces, which defines the state of mechanical reference. The force field at the surface of the gel is calculated from the measurement of the particle displacements using the procedure described above. It should be stressed that we do not apply any mathematical treatment, such as interpolation of particle positions, interpolation of the field of forces or smoothing of the computed force field. Moreover, no \textit{a priori} assumptions are performed concerning the points of application of the forces: they are computed on a quadratic grid with a $3\ \mu$m mesh size.  The spatial resolution of the force field is solely determined by particle concentration and by the accuracy of the measurement of their displacements. In particular along the $z$ direction, this allows for the precise measurement of all the components of the force field at the surface of the gel.

\begin{figure}[h!]
      \begin{center}
              \includegraphics[width=0.7\textwidth]{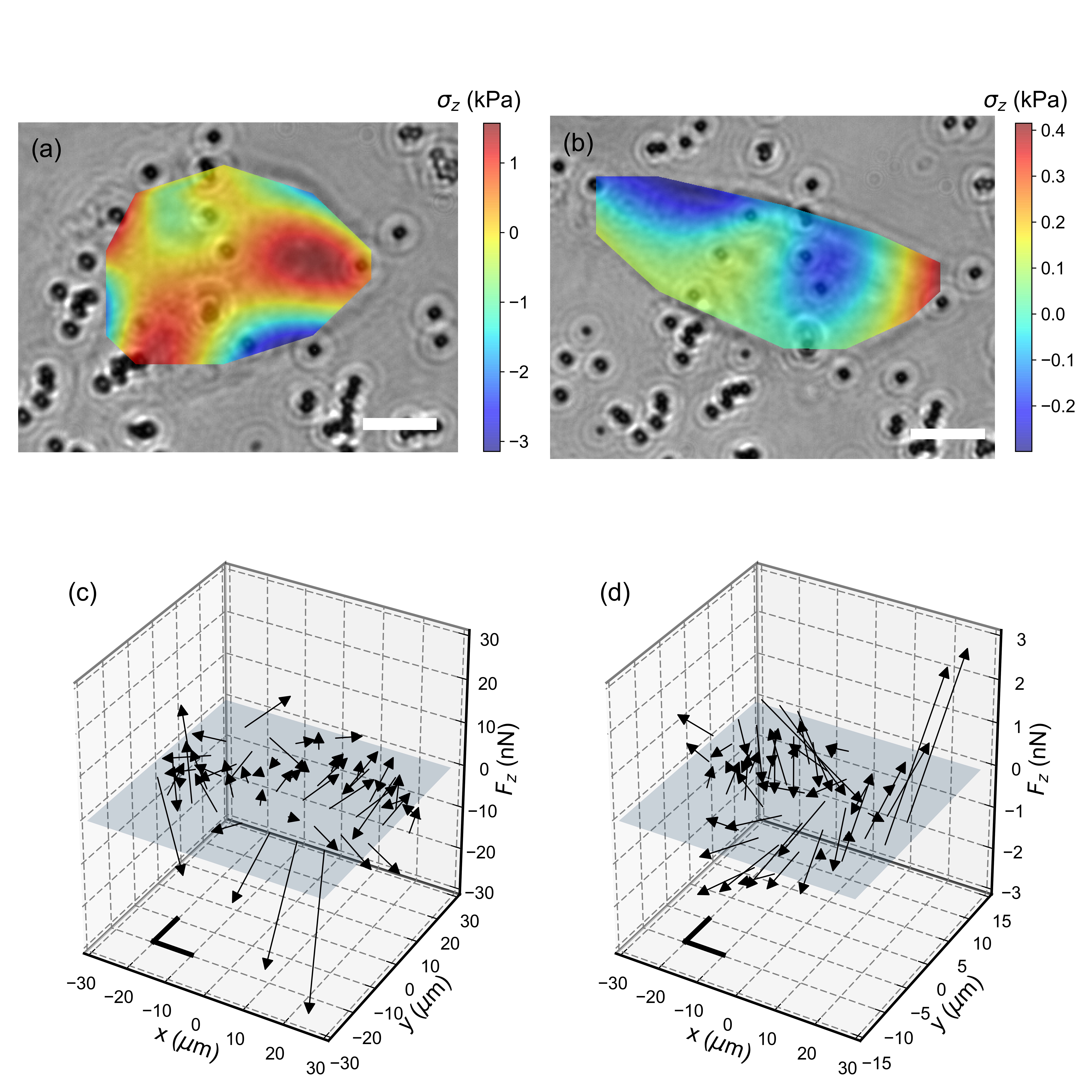}
            \end{center}
\caption{\textbf{(a)} and \textbf{(b)} Bright field images of a round \textbf{(a)} and elongated \textbf{(b)} SW480 cell, superimposed with a color plot of the exerted normal stresses exerted by the cell. Positive stresses corresponds to cells pulling the substrate. Scale bars, $10\ \mu$m. \textbf{(c)} and \textbf{(d)} Three-dimensional plot of the forces exerted by a round \textbf{(c)} and elongated \textbf{(d)}  cell onto the substrate.  The thick black segments represents forces along the $x$ and $y$ directions equal to $10$ nN \textbf{(c)} and $1$ nN \textbf{(d)}.  }
\label{fig:forcefield}
\end{figure}

We have found that cells exhibit two phenotypes. First, cells may exhibit round shapes. In this case, cells exert large forces around their periphery; this results in simultaneous pushing and pulling of the substrates into the adhesion region. (Fig.~\ref{fig:forcefield} \textbf{(a), (c)}). The applied pressure in the center of the cell is smaller than at its boundary so that the overall sum of the forces is null. More precisely, computing the normalized sum of the forces over the entire force field, 
\begin{equation}
\delta = \frac{| \sum_{k=1}^{N} \vec{F}(\vec{r}_k)|}{ \sum_{k=1}^{N}|\vec{F}(\vec{r}_k)|}
\label{delta}
\end{equation}
\noindent, we found that, the average values of $\delta$ over the round cell shapes is $\delta = 0.013$. \\

Second, cells can adopt an elongated geometry (Fig.~\ref{fig:forcefield} \textbf{(a), (c)}). Here, two stress peaks  are  observed on opposite poles of the cell. One of the force peaks is directed into the substrate whereas the other is directed out of the substrate (Fig.~\ref{fig:forcefield} \textbf{(b)}). The amplitude of these force peaks is such that the sum of all forces vanishes: $\delta = 0.085$. In other words, the cell pushes the gel at one of its poles and pulls at the opposite extremity.  \\

\begin{figure}[h!]
      \begin{center}
              \includegraphics[width=0.7\textwidth]{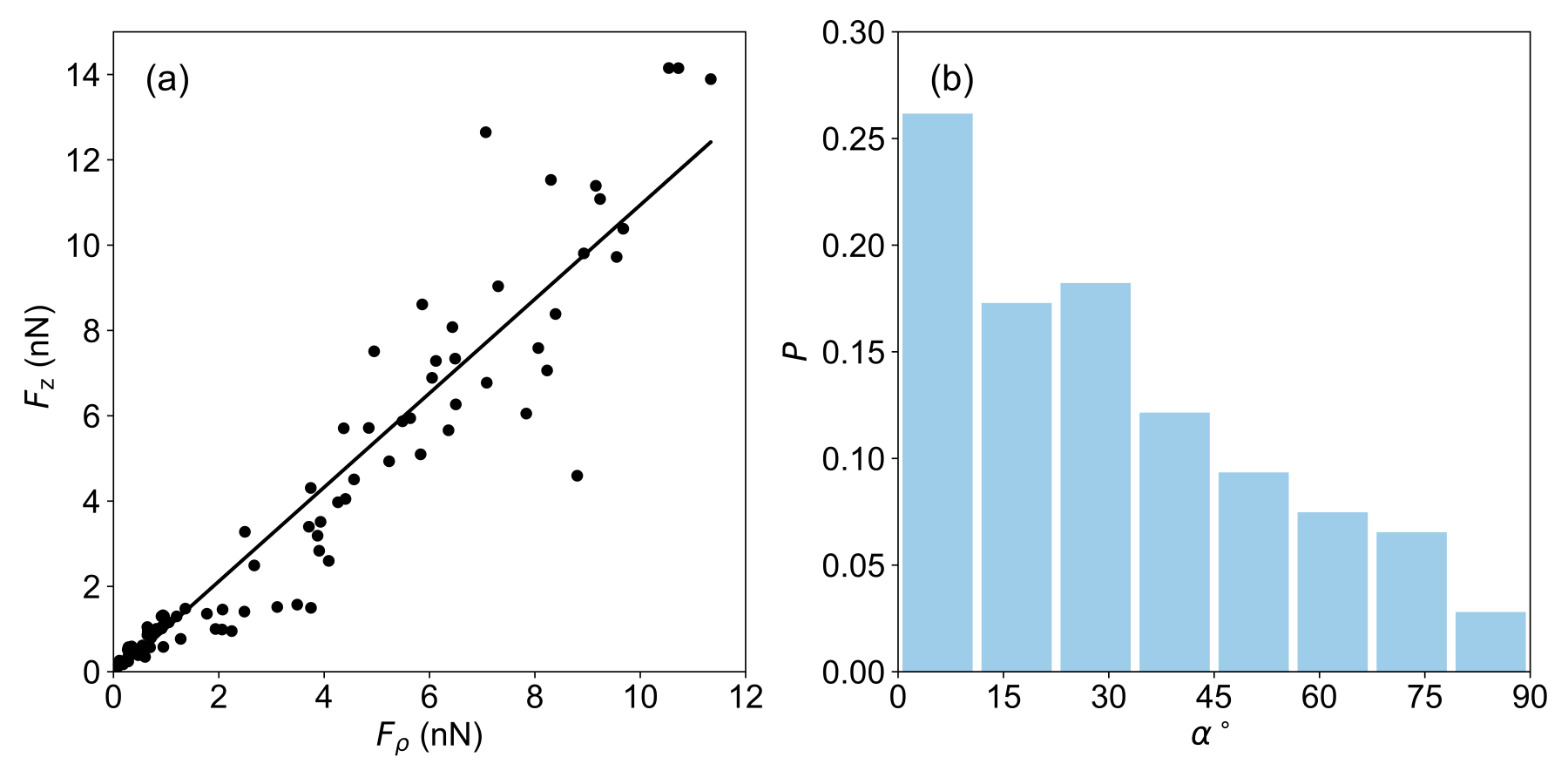}
            \end{center}
\caption{\textbf{(a)} Normal force $F_z$ as a function of the component of the force in the $xy$ plane, $F_\rho$. Both elongated and round shapes are taken into account. Each point corresponds to the average value of force vectors over an entire field of forces, taken over $170$ measurements, of both elongated and round cell shapes. Black line is a linear adjustement of the data. The slope is $1.09$. \textbf{(b)} Distribution of the values of the angle $\alpha$ between the main axis of the cell elongation and the major dipole axis of the force field. Major dipole axes were computed with  the procedure, described in the text; cell elongation axis are determined by an ellipsoidal adjustement of the cellular boundary. $210$ images were analyzed.}

\label{fig:dipole}
\end{figure}

Interestingly, for both shapes, the $z$-component of the forces exerted by the cell onto the substrate are of the same order of magnitude as the shear forces. If one plots the amplitude of the normal forces as the  function of tangential forces, both shapes taken into account (Fig.~\ref{fig:dipole} \textbf{(a)}), a linear relation is obtained (slope of $1.09$). This indicates that tangential and normal forces are comparable in magnitude. A similar behavior has been reported for \textit{Dictyostelium} Cells~\cite{delanoe-ayari_4d_2010}, although the normal component of the force was slightly smaller than the tangential one (slope $0.72$), and for mammalian cells (fibroblats)~\cite{Maskarinec2009}. 

For an elongated cell, the normal component of the applied force has a dipolar behavior. A similar analysis can be applied to the tangential component of the force field. Following Tanimoto's approach~\cite{tanimoto_simple_2014}, let us consider the first non-zero moment of a multipolar expansion of the force field matrix,  the dipolar term:

\begin{equation}
M_{ij}=\sum_{k=1}^{N} x_i(\vec{r}_k)F_j(\vec{r}_k)
\label{forcedipole}
\end{equation}

\noindent where the sum is taken over all $N$ positions of the force vectors underneath the cell; $x_i$ and $F_j$ are the $i$th and $j$th components of the positions with respect to  the cell center and the measured force, respectively ($i=1 (2)$ designates $x (y)$ axis).  The total torque vanishes and the matrix $M_{ij}$ is thus symmetric and diagonalizable. 

Let us define the major dipole as the eigenvector with the largest eigenvalue. The associated eigenvalue is negative, corresponding to a contractile behavior of the cell along this direction. For elongated cells, the contraction axis is correlated with the shape anisotropy of the cell itself.  For each of these force fields, defining the angle $\alpha$ between the cell elongation axis and the major dipole axis (Fig.~\ref{fig:dipole} \textbf{(b)}), we observe that the histogram exhibits a maximum for small $\alpha$ values (Fig.~\ref{fig:dipole} \textbf{(b)}). This indicates that the force axis lies within the the long axis of the cell.

The coexistence of two  morphologies  has already been reported for  SW480 cells. Their mechanical properties have been studied~\cite{palmieri_mechanical_2015}: the Young modulus of round shape cells ($500$ Pa) was found to be  smaller than that of elongated cells, and the adhesion of cells onto an Atomic Force Microscope cantilever was found to be independent of the cellular shape. Our results constitute the first study of the adhesion pattern for each cellular shape and show that, although the elastic properties of eongated and round SW480 cells are similar, their adhesion patterns strongly differ.

\section*{Conclusion}

Here, we have introduced a new TFM approach, which uses non-fluorescent particles. This technique allows to track micrometer-sized particles (along all three directions) with a localization accuracy that cannot be achieved using state-of-the art TFM (fluorescent-based) setups. This low tracking noise ($\sim$ nm) allows to successfully recover force maps, including the normal component of the forces, at low spatial resolutions (2D density smaller than $0.1$ particles per $\mu$m$^2$. This technique is still new and further improvements are expected. Sub-nm localization accuracies could be obtained using a faster camera (capable of nearly kHz acquisition at full frame) and higher power light sources such as superluminescent diodes  (simply because averaging decreases tracking noise~\cite{huhle_camera-based_2015}), which also offer excellent image quality~\cite{dulin_efficient_2014}. Here, the typical extension of diffraction patterns could be controlled along both the $x$ and $y$ axis by using spatial filtering and select low frequency components and along the $z$ axis by confining the beads in one plane only. Assuming an intensity of $1$ mW, and an illuminated region of $100$X$100$ $\mu$m$^2$, the flux is $10$~W \textit{per}  cm$^2$ (which is sufficient to observe well-defined diffraction patterns at $0.1$~kHz~\cite{cavatore_microscopie_2011}), the typical dose is about $1$~$J$ \textit{per} cm$^2$ for 100 images in $1$~s. Repeating the acquisition every minute for 10 hours, the total dose is 600~J per cm$^2$ and so would not damage cells (using a wavelength larger than 600~nm)~\cite{waldchen2015}. Obviously, fluorescence imaging is also capable to track nanobeads with nm accuracy~\cite{Gardini2015}. This, however, requires large integration times (averaging  $10$  images over $150$ ms allows for an accuracy about $1$ to $2$~nm in all three directions, \textit{i.e.} roughly an order of magnitude higher than what is expected using non-fluorescent measurements) and much higher intensities (about $100$ times larger), which would result in potential photo-damage  and photo-bleaching effects. 
Finally, it remains also possible, using Mie scattering theory, to track particles when diffraction patterns overlap~\cite{fung_imaging_2012}, and to reach very large particle volume fractions. We believe that our new approach should stimulate new theoretical investigations in order to optimize both volume fractions and accuracy parameters.

\section*{Materials and methods}
\label{sec:matmeth}

\subsection*{TFM Setup}
We use a home-built microscope. Bead images (8 bits) were acquired with a 2048 x 2048 pixels CMOS camera (acA2040-25gm, Basler) that has a saturation capacity of 11.9 ke- and a frame rate of 25. An oil-immersion objective (100X, NA 1.25, Zeiss) was mounted on a piezoelectric flexure objective scanner (P-721, Physik Instrumente) and used to image the gel at different positions (along the optical axis). A lens in front of the camera  sets the magnification to about 50. To maintain a relatively low spatial coherence (to about $6$ micrometers), we use a Light Emitting Diode (M595L3, Thorlabs) and a band pass filter (FF01-697/75-25-D, Semrock). To minimize temperature gradients, the stainless steel microscope stage is thermally isolated from the optical table with ceramic legs. Experiments were performed at T=37 $^\circ$C (TempController 2000-2, Pecon GmbH) under a 5 percent  Carbon Dioxide atmosphere (CO2-Controller 2000, Pecon GmbH).

\subsection*{Image Acquisition}
Unless specified, 20 images were acquired at  20 Hz every minute and then averaged. To correct for the difference in index of refractions between oil and water, the $z$ positions were multiplied by a factor 0.82$\pm$ 0.01~\cite{hell_1993_aberrations}. Note that this experimental value (obtained by measuring the thickness of different flow cells) deviates from the ratio of indexes (1.33/1.515=0.88, assuming a low NA) but is in agreement with a model proposed by  Visser~\cite{visser_1992_refractive}.

\subsection*{Preparation of Activated Coverslips}
We used 35mm glass bottomed Cell Culture Dishes (500027, Porvair). To covalently attach the polyacrylamide gel onto glass, we used a a similar procedure as in \cite{fischer_stiffness-controlled_2012}. Glass surfaces were cleaned with NaOH, incubated with a 0.5$\%$ EtOH solution of 3-Aminopropyltriethoxysilane (440140, SIGMA) and then immersed in a 0.5$\%$ Glutaraldehyde (G6257, SIGMA). Intensive rinsing with either H20 or EtOH was performed between all steps.

\subsection*{Polyacrylamide gel fabrication} 
Gels with $80$ micrometer thickness were polymerized onto functionalized glass using the following protocol \cite{fischer_stiffness-controlled_2012}; A solution of acrylamide (5\%; 1610142, Bio-Rad) and bis-acrylamide (0.05\%; 1610140, Bio-Rad) was mixed with $1$ micrometer diameter polystyrene particles (07310, Polysciences) to yield a gel with a Young modulus of $E=0.45$ kPa~\cite{fischer_stiffness-controlled_2012}. The concentration of particles was adjusted to obtain a volume fraction of about 0.1\%. Polymerization was initiated by Ammonium Persulfate (A3678, SIGMA) and Tetramethylethylenediamine (T9281, SIGMA). After complete polymerization (about 30 minutes), Collagen I (0.2 mg/ml in Acetic Acid; A10483, Life Technologies) was cross-linked to the gel surface using Sulfo-SANPAH (1 mM;  BC38, G-Biosciences). Photoactivation was performed with UV and cross-linking was done overnight. The cell culture dishes were then stored in Phosphate-buffered Saline buffer (79382, SIGMA) at 4 $^\circ$C.

\subsection*{Cell Culture}
SW480 cells were obtained from ATCC and grown in DMEM (Dulbecco's modified Eagle's medium; Life Technology) with 10\% fetal bovine serum (Life technology, Germany) at  37 $^\circ$C  in a humidified atmosphere and 5\% CO2. Mycoplasma contamination has been tested negatively using PlasmoTest (Invivo gene). Cells were seeded on the gel-
covered slides at a concentration of 50 000 cells/ml to avoid confluency and allow individual cell measurment. Cells were maintained at 37 $^\circ$C and 5\% CO2 during measurements using a dedicated chamber.  Note that our procedure is not compatible with some protocols used in macroscopic cell culture but is similar to protocols used in Microfluidics~\cite{Halldorsson2015}).

\subsection*{Data Availability} 
The datasets generated during and/or analysed during the current study are available from the corresponding author on reasonable request.

\section*{Author contributions}
CG, PH and WG conceived the experiments. SM  performed the experiments. PH, SM and WG analyzed data. NB designed and built the setup. SM and PH performed to the theoretical analysis.  SM and WG wrote the code to control the setup and track particles. PH, SM and WG wrote the paper. All authors reviewed the manuscript.

\section*{Competing financial interests}
The authors declare no competing financial interests.


\begin{thebibliography}{10}
\expandafter\ifx\csname url\endcsname\relax
  \def\url#1{\texttt{#1}}\fi
\expandafter\ifx\csname urlprefix\endcsname\relax\def\urlprefix{URL }\fi
\providecommand{\bibinfo}[2]{#2}
\providecommand{\eprint}[2][]{\url{#2}}

\bibitem{Wolfenson2009}
\bibinfo{author}{Wolfenson, H.}, \bibinfo{author}{Henis, Y.~I.},
  \bibinfo{author}{Geiger, B.} \& \bibinfo{author}{Bershadsky, A.~D.}
\newblock \bibinfo{title}{The heel and toe of the cell's foot: {{A}}
  multifaceted approach for understanding the structure and dynamics of focal
  adhesions}.
\newblock \emph{\bibinfo{journal}{Cell motility and the cytoskeleton}}
  \textbf{\bibinfo{volume}{66}}, \bibinfo{pages}{1017--1029}
  (\bibinfo{year}{2009}).

\bibitem{Cukierman2001a}
\bibinfo{author}{Cukierman, E.}, \bibinfo{author}{Pankov, R.},
  \bibinfo{author}{Stevens, D.~R.} \& \bibinfo{author}{Yamada, K.~M.}
\newblock \bibinfo{title}{Taking {{Cell-Matrix Adhesions}} to the {{Third
  Dimension}}}.
\newblock \emph{\bibinfo{journal}{Science}} \textbf{\bibinfo{volume}{294}},
  \bibinfo{pages}{1708--1712} (\bibinfo{year}{2001}).

\bibitem{Discher2005a}
\bibinfo{author}{Discher, D.~E.}
\newblock \bibinfo{title}{Tissue {{Cells Feel}} and {{Respond}} to the
  {{Stiffness}} of {{Their Substrate}}}.
\newblock \emph{\bibinfo{journal}{Science}} \textbf{\bibinfo{volume}{310}},
  \bibinfo{pages}{1139--1143} (\bibinfo{year}{2005}).

\bibitem{agus_physical_2013}
\bibinfo{author}{Agus, D.~B.} \emph{et~al.}
\newblock \bibinfo{title}{A physical sciences network characterization of
  non-tumorigenic and metastatic cells}.
\newblock \emph{\bibinfo{journal}{Scientific Reports}}
  \textbf{\bibinfo{volume}{3}} (\bibinfo{year}{2013}).

\bibitem{hochmuth_micropipette_2000}
\bibinfo{author}{Hochmuth, R.~M.}
\newblock \bibinfo{title}{Micropipette aspiration of living cells}.
\newblock \emph{\bibinfo{journal}{Journal of biomechanics}}
  \textbf{\bibinfo{volume}{33}}, \bibinfo{pages}{15--22}
  (\bibinfo{year}{2000}).

\bibitem{lu_microfluidic_2004}
\bibinfo{author}{Lu, H.} \emph{et~al.}
\newblock \bibinfo{title}{Microfluidic shear devices for quantitative analysis
  of cell adhesion}.
\newblock \emph{\bibinfo{journal}{Analytical chemistry}}
  \textbf{\bibinfo{volume}{76}}, \bibinfo{pages}{5257--5264}
  (\bibinfo{year}{2004}).

\bibitem{Harris1980}
\bibinfo{author}{Harris, A.}, \bibinfo{author}{Wild, P.} \&
  \bibinfo{author}{Stopak, D.}
\newblock \bibinfo{title}{Silicone rubber substrata: a new wrinkle in the study
  of cell locomotion}.
\newblock \emph{\bibinfo{journal}{Science}} \textbf{\bibinfo{volume}{208}},
  \bibinfo{pages}{177--179} (\bibinfo{year}{1980}).

\bibitem{Harris1981}
\bibinfo{author}{Harris, A.~K.}, \bibinfo{author}{Stopak, D.} \&
  \bibinfo{author}{Wild, P.}
\newblock \bibinfo{title}{Fibroblast traction as a mechanism for collagen
  morphogenesis}.
\newblock \emph{\bibinfo{journal}{Nature}} \textbf{\bibinfo{volume}{290}},
  \bibinfo{pages}{249--251} (\bibinfo{year}{1981}).

\bibitem{Galbraith1997}
\bibinfo{author}{Galbraith, C.~G.} \& \bibinfo{author}{Sheetz, M.~P.}
\newblock \bibinfo{title}{A micromachined device provides a new bend on
  fibroblast traction\,forces}.
\newblock \emph{\bibinfo{journal}{Proceedings of the National Academy of
  Sciences of the United States of America}} \textbf{\bibinfo{volume}{94}},
  \bibinfo{pages}{9114--9118} (\bibinfo{year}{1997}).

\bibitem{tan_cells_2003}
\bibinfo{author}{Tan, J.~L.} \emph{et~al.}
\newblock \bibinfo{title}{Cells lying on a bed of microneedles: {An} approach
  to isolate mechanical force}.
\newblock \emph{\bibinfo{journal}{Proceedings of the National Academy of
  Sciences}} \textbf{\bibinfo{volume}{100}}, \bibinfo{pages}{1484--1489}
  (\bibinfo{year}{2003}).

\bibitem{du_roure_force_2005}
\bibinfo{author}{du~Roure, O.} \emph{et~al.}
\newblock \bibinfo{title}{Force mapping in epithelial cell migration}.
\newblock \emph{\bibinfo{journal}{Proceedings of the National Academy of
  Sciences}} \textbf{\bibinfo{volume}{102}}, \bibinfo{pages}{2390--2395}
  (\bibinfo{year}{2005}).

\bibitem{saez_rigidity-driven_2007}
\bibinfo{author}{Saez, A.}, \bibinfo{author}{Ghibaudo, M.},
  \bibinfo{author}{Buguin, A.}, \bibinfo{author}{Silberzan, P.} \&
  \bibinfo{author}{Ladoux, B.}
\newblock \bibinfo{title}{Rigidity-driven growth and migration of epithelial
  cells on microstructured anisotropic substrates}.
\newblock \emph{\bibinfo{journal}{Proceedings of the National Academy of
  Sciences}} \textbf{\bibinfo{volume}{104}}, \bibinfo{pages}{8281--8286}
  (\bibinfo{year}{2007}).

\bibitem{tee_cell_2011}
\bibinfo{author}{Tee, S.-Y.}, \bibinfo{author}{Fu, J.}, \bibinfo{author}{Chen,
  C.~S.} \& \bibinfo{author}{Janmey, P.~A.}
\newblock \bibinfo{title}{Cell shape and substrate rigidity both regulate cell
  stiffness}.
\newblock \emph{\bibinfo{journal}{Biophysical Journal}}
  \textbf{\bibinfo{volume}{100}}, \bibinfo{pages}{L25--L27}
  (\bibinfo{year}{2011}).

\bibitem{Dembo1999}
\bibinfo{author}{Dembo, M.} \& \bibinfo{author}{Wang, Y.-L.}
\newblock \bibinfo{title}{Stresses at the {{Cell}}-to-{{Substrate Interface}}
  during {{Locomotion}} of {{Fibroblasts}}}.
\newblock \emph{\bibinfo{journal}{Biophysical Journal}}
  \textbf{\bibinfo{volume}{76}}, \bibinfo{pages}{2307--2316}
  (\bibinfo{year}{1999}).

\bibitem{del_alamo_three-dimensional_2013}
\bibinfo{author}{del Ãlamo, J.~C.} \emph{et~al.}
\newblock \bibinfo{title}{Three-dimensional quantification of cellular traction
  forces and mechanosensing of thin substrata by fourier traction force
  microscopy}.
\newblock \emph{\bibinfo{journal}{{PLoS} {ONE}}} \textbf{\bibinfo{volume}{8}},
  \bibinfo{pages}{e69850} (\bibinfo{year}{2013}).

\bibitem{Butler2002}
\bibinfo{author}{Butler, J.~P.}, \bibinfo{author}{Toli{\'c}-N\o{}rrelykke,
  I.~M.}, \bibinfo{author}{Fabry, B.} \& \bibinfo{author}{Fredberg, J.~J.}
\newblock \bibinfo{title}{Traction fields, moments, and strain energy that
  cells exert on their surroundings}.
\newblock \emph{\bibinfo{journal}{American Journal of Physiology - Cell
  Physiology}} \textbf{\bibinfo{volume}{282}}, \bibinfo{pages}{C595--C605}
  (\bibinfo{year}{2002}).

\bibitem{Schwarz2002}
\bibinfo{author}{Schwarz, U.~S.} \emph{et~al.}
\newblock \bibinfo{title}{Calculation of forces at focal adhesions from elastic
  substrate data: the effect of localized force and the need for
  regularization.}
\newblock \emph{\bibinfo{journal}{Biophysical Journal}}
  \textbf{\bibinfo{volume}{83}}, \bibinfo{pages}{1380--1394}
  (\bibinfo{year}{2002}).

\bibitem{Sabass2008}
\bibinfo{author}{Sabass, B.}, \bibinfo{author}{Gardel, M.~L.},
  \bibinfo{author}{Waterman, C.~M.} \& \bibinfo{author}{Schwarz, U.~S.}
\newblock \bibinfo{title}{High {{Resolution Traction Force Microscopy Based}}
  on {{Experimental}} and {{Computational Advances}}}.
\newblock \emph{\bibinfo{journal}{Biophysical Journal}}
  \textbf{\bibinfo{volume}{94}}, \bibinfo{pages}{207--220}
  (\bibinfo{year}{2008}).

\bibitem{zundel_factors_2017}
\bibinfo{author}{Z\"undel, M.}, \bibinfo{author}{Ehret, A.~E.} \&
  \bibinfo{author}{Mazza, E.}
\newblock \bibinfo{title}{Factors influencing the determination of cell
  traction forces}.
\newblock \emph{\bibinfo{journal}{PLOS ONE}} \textbf{\bibinfo{volume}{12}},
  \bibinfo{pages}{e0172927} (\bibinfo{year}{2017}).

\bibitem{bergert_confocal_2016}
\bibinfo{author}{Bergert, M.} \emph{et~al.}
\newblock \bibinfo{title}{Confocal reference free traction force microscopy}.
\newblock \emph{\bibinfo{journal}{Nature Communications}}
  \textbf{\bibinfo{volume}{7}}, \bibinfo{pages}{12814} (\bibinfo{year}{2016}).

\bibitem{colin-york_super-resolved_2016}
\bibinfo{author}{Colin-York, H.} \emph{et~al.}
\newblock \bibinfo{title}{Super-resolved traction force microscopy ({STFM})}.
\newblock \emph{\bibinfo{journal}{Nano Letters}} \textbf{\bibinfo{volume}{16}},
  \bibinfo{pages}{2633--2638} (\bibinfo{year}{2016}).

\bibitem{colin-york2018}
\bibinfo{author}{Colin-York, H.} \& \bibinfo{author}{Fritzsche, M.}
\newblock \bibinfo{title}{The future of traction force microscopy}.
\newblock \emph{\bibinfo{journal}{Current Opinion in Biomedical Engineering}}
  \textbf{\bibinfo{volume}{5}}, \bibinfo{pages}{1--5} (\bibinfo{year}{2018}).

\bibitem{huang_three-dimensional_2008}
\bibinfo{author}{Huang, B.}, \bibinfo{author}{Wang, W.},
  \bibinfo{author}{Bates, M.} \& \bibinfo{author}{Zhuang, X.}
\newblock \bibinfo{title}{Three-{Dimensional} {Super}-{Resolution} {Imaging} by
  {Stochastic} {Optical} {Reconstruction} {Microscopy}}.
\newblock \emph{\bibinfo{journal}{Science}} \textbf{\bibinfo{volume}{319}},
  \bibinfo{pages}{810--813} (\bibinfo{year}{2008}).

\bibitem{waldchen2015}
\bibinfo{author}{W\"aldchen, A.}, \bibinfo{author}{Lehmann, J.},
  \bibinfo{author}{Klein, T.}, \bibinfo{author}{can~de Linde, S.} \&
  \bibinfo{author}{Sauer, M.}
\newblock \bibinfo{title}{Light-induced cell damage in live-cell
  super-resolution microscopy}.
\newblock \emph{\bibinfo{journal}{Scientific Reports}}
  \textbf{\bibinfo{volume}{5}} (\bibinfo{year}{2015}).

\bibitem{wirtz2011}
\bibinfo{author}{Wirtz, D.}, \bibinfo{author}{Konstantopoulos, K.} \&
  \bibinfo{author}{Searson, P.}
\newblock \bibinfo{title}{The physics of cancer: the role of physical
  interactions and mechanical forces in metastasis}.
\newblock \emph{\bibinfo{journal}{Nature Reviews. Cancer.}}
  \textbf{\bibinfo{volume}{11}}, \bibinfo{pages}{512--522}
  (\bibinfo{year}{2011}).

\bibitem{gosse_magnetic_2002}
\bibinfo{author}{Gosse, C.} \& \bibinfo{author}{Croquette, V.}
\newblock \bibinfo{title}{Magnetic {Tweezers}: {Micromanipulation} and {Force}
  {Measurement} at the {Molecular} {Level}}.
\newblock \emph{\bibinfo{journal}{Biophysical Journal}}
  \textbf{\bibinfo{volume}{82}}, \bibinfo{pages}{3314--3329}
  (\bibinfo{year}{2002}).

\bibitem{delanoe-ayari_4d_2010}
\bibinfo{author}{Delano\"e-Ayari, H.}, \bibinfo{author}{Rieu, J.~P.} \&
  \bibinfo{author}{Sano, M.}
\newblock \bibinfo{title}{4d traction force microscopy reveals asymmetric
  cortical forces in migrating \textit{Dictyostelium} cells}.
\newblock \emph{\bibinfo{journal}{Physical Review Letters}}
  \textbf{\bibinfo{volume}{105}} (\bibinfo{year}{2010}).

\bibitem{lionnet_single-molecule_2012}
\bibinfo{author}{Lionnet, T.} \emph{et~al.}
\newblock \bibinfo{title}{Single-molecule studies using magnetic traps}.
\newblock \emph{\bibinfo{journal}{Cold Spring Harbor Protocols}}
  \textbf{\bibinfo{volume}{2012}}, \bibinfo{pages}{067488}
  (\bibinfo{year}{2012}).

\bibitem{zhang_three-dimensional_2008}
\bibinfo{author}{Zhang, Z.} \& \bibinfo{author}{Menq, C.-H.}
\newblock \bibinfo{title}{Three-dimensional particle tracking with subnanometer
  resolution using off-focus images}.
\newblock \emph{\bibinfo{journal}{Applied optics}}
  \textbf{\bibinfo{volume}{47}}, \bibinfo{pages}{2361--2370}
  (\bibinfo{year}{2008}).

\bibitem{vanloenhout_non-bias-limited_2012}
\bibinfo{author}{van Loenhout, M.~T.}, \bibinfo{author}{Kerssemakers, J.~W.},
  \bibinfo{author}{De~Vlaminck, I.} \& \bibinfo{author}{Dekker, C.}
\newblock \bibinfo{title}{Non-bias-limited tracking of spherical particles,
  enabling nanometer resolution at low magnification}.
\newblock \emph{\bibinfo{journal}{Biophysical Journal}}
  \textbf{\bibinfo{volume}{102}}, \bibinfo{pages}{2362--2371}
  (\bibinfo{year}{2012}).

\bibitem{cavatore_microscopie_2011}
\bibinfo{author}{Cavatore, E.}
\newblock \emph{\bibinfo{title}{{Optical Microscopy applied to
  micro-manipulation by high-resolution magnetic tweezers and to visualization
  of metal nano-objects}}}.
\newblock \bibinfo{type}{Phd thesis}, \bibinfo{school}{{Universit{\'e} Pierre
  et Marie Curie - Paris VI}} (\bibinfo{year}{2011}).

\bibitem{cnossen_optimized_2014}
\bibinfo{author}{Cnossen, J.~P.}, \bibinfo{author}{Dulin, D.} \&
  \bibinfo{author}{Dekker, N.~H.}
\newblock \bibinfo{title}{An optimized software framework for real-time,
  high-throughput tracking of spherical beads}.
\newblock \emph{\bibinfo{journal}{Review of Scientific Instruments}}
  \textbf{\bibinfo{volume}{85}}, \bibinfo{pages}{103712}
  (\bibinfo{year}{2014}).

\bibitem{allan_statistics_1966}
\bibinfo{author}{Allan, D.}
\newblock \bibinfo{title}{Statistics of atomic frequency standards}.
\newblock \emph{\bibinfo{journal}{Proceedings of the IEEE}}
  \textbf{\bibinfo{volume}{54}}, \bibinfo{pages}{221--230}
  (\bibinfo{year}{1966}).

\bibitem{czerwinski_quantifying_2009}
\bibinfo{author}{Czerwinski, F.}, \bibinfo{author}{Richardson, A.~C.} \&
  \bibinfo{author}{Oddershede, L.~B.}
\newblock \bibinfo{title}{Quantifying {Noise} in {Optical} {Tweezers} by
  {Allan} {Variance}}.
\newblock \emph{\bibinfo{journal}{Optics Express}}
  \textbf{\bibinfo{volume}{17}}, \bibinfo{pages}{13255} (\bibinfo{year}{2009}).

\bibitem{dulin_efficient_2014}
\bibinfo{author}{Dulin, D.}, \bibinfo{author}{Barland, S.},
  \bibinfo{author}{Hachair, X.} \& \bibinfo{author}{Pedaci, F.}
\newblock \bibinfo{title}{Efficient {Illumination} for {Microsecond} {Tracking}
  {Microscopy}}.
\newblock \emph{\bibinfo{journal}{PLoS ONE}} \textbf{\bibinfo{volume}{9}},
  \bibinfo{pages}{e107335} (\bibinfo{year}{2014}).

\bibitem{vlaminck2011}
\bibinfo{author}{De~Vlaminck, I.} \emph{et~al.}
\newblock \bibinfo{title}{Highly parallel magnetic tweezers by targeted dna
  tethering}.
\newblock \emph{\bibinfo{journal}{Nanoletters}} \textbf{\bibinfo{volume}{11}},
  \bibinfo{pages}{5489--5493} (\bibinfo{year}{2011}).

\bibitem{LANDAU1976}
\bibinfo{author}{Landau, L.} \& \bibinfo{author}{Lifshitz, E.}
\newblock \bibinfo{title}{{The} {equations} {of} {motion}}.
\newblock In \emph{\bibinfo{booktitle}{Mechanics}}, \bibinfo{pages}{1--12}
  (\bibinfo{publisher}{Elsevier}, \bibinfo{year}{1976}).

\bibitem{alvarez-gonzalez_two-layer_2017}
\bibinfo{author}{Alvarez-Gonzalez, B.} \emph{et~al.}
\newblock \bibinfo{title}{Two-{Layer} {Elastographic} 3-{D} {Traction} {Force}
  {Microscopy}}.
\newblock \emph{\bibinfo{journal}{Scientific Reports}}
  \textbf{\bibinfo{volume}{7}}, \bibinfo{pages}{39315} (\bibinfo{year}{2017}).

\bibitem{kirsch2011introduction}
\bibinfo{author}{Kirsch, A.}
\newblock \emph{\bibinfo{title}{An introduction to the mathematical theory of
  inverse problems}}, vol. \bibinfo{volume}{120} (\bibinfo{publisher}{Springer
  Science \& Business Media}, \bibinfo{year}{2011}).

\bibitem{tikhonov1977solutions}
\bibinfo{author}{Tikhonov, A.~N.} \& \bibinfo{author}{Arsenin, V.~Y.}
\newblock \emph{\bibinfo{title}{Solutions of ill-posed problems}}
  (\bibinfo{address}{John Wiley \& Sons, New York}, \bibinfo{year}{1977}).

\bibitem{hansen_l-curve_1999}
\bibinfo{author}{Hansen, P.~C.}
\newblock \emph{\bibinfo{title}{The L-curve and its use in the numerical
  treatment of inverse problems}} (\bibinfo{publisher}{{IMM}, Department of
  Mathematical Modelling, Technical University of Denmark},
  \bibinfo{year}{1999}).

\bibitem{Schwarz2015}
\bibinfo{author}{Schwarz, U.~S.} \& \bibinfo{author}{Soin{\'e}, J. R.~D.}
\newblock \bibinfo{title}{Traction force microscopy on soft elastic substrates:
  {{A}} guide to recent computational advances}.
\newblock \emph{\bibinfo{journal}{Biochimica et Biophysica Acta (BBA) -
  Molecular Cell Research}} \textbf{\bibinfo{volume}{1853}},
  \bibinfo{pages}{3095--3104} (\bibinfo{year}{2015}).

\bibitem{hansen_2007_regularization}
\bibinfo{author}{Hansen, P.~C.}
\newblock \bibinfo{title}{Regularization tools version 4.0 for matlab 7.3}.
\newblock \emph{\bibinfo{journal}{Numerical algorithms}}
  \textbf{\bibinfo{volume}{46}}, \bibinfo{pages}{189--194}
  (\bibinfo{year}{2007}).

\bibitem{Maskarinec2009}
\bibinfo{author}{Maskarinec, S.~A.}, \bibinfo{author}{Franck, C.},
  \bibinfo{author}{Tirrell, D.~A.} \& \bibinfo{author}{Ravichandran, G.}
\newblock \bibinfo{title}{Quantifying cellular traction forces in three
  dimensions}.
\newblock \emph{\bibinfo{journal}{Proceedings of the National Academy of
  Sciences}} \textbf{\bibinfo{volume}{106}}, \bibinfo{pages}{22108--22113}
  (\bibinfo{year}{2009}).

\bibitem{tanimoto_simple_2014}
\bibinfo{author}{Tanimoto, H.} \& \bibinfo{author}{Sano, M.}
\newblock \bibinfo{title}{A simple force-motion relation for migrating cells
  revealed by multipole analysis of traction stress}.
\newblock \emph{\bibinfo{journal}{Biophysical Journal}}
  \textbf{\bibinfo{volume}{106}}, \bibinfo{pages}{16--25}
  (\bibinfo{year}{2014}).

\bibitem{palmieri_mechanical_2015}
\bibinfo{author}{Palmieri, V.} \emph{et~al.}
\newblock \bibinfo{title}{Mechanical and structural comparison between primary
  tumor and lymph node metastasis cells in colorectal cancer}.
\newblock \emph{\bibinfo{journal}{Soft Matter}} \textbf{\bibinfo{volume}{11}},
  \bibinfo{pages}{5719--5726} (\bibinfo{year}{2015}).

\bibitem{huhle_camera-based_2015}
\bibinfo{author}{Huhle, A.} \emph{et~al.}
\newblock \bibinfo{title}{Camera-based three-dimensional real-time particle
  tracking at {kHz} rates and angstrom accuracy}.
\newblock \emph{\bibinfo{journal}{Nature Communications}}
  \textbf{\bibinfo{volume}{6}}, \bibinfo{pages}{5885} (\bibinfo{year}{2015}).

\bibitem{Gardini2015}
\bibinfo{author}{Gardini, L.}, \bibinfo{author}{Capitanio, M.} \&
  \bibinfo{author}{Pavone, F.~S.}
\newblock \bibinfo{title}{{{3D}} tracking of single nanoparticles and quantum
  dots in living cells by out-of-focus imaging with diffraction pattern
  recognition}.
\newblock \emph{\bibinfo{journal}{Scientific Reports}}
  \textbf{\bibinfo{volume}{5}}, \bibinfo{pages}{16088} (\bibinfo{year}{2015}).

\bibitem{fung_imaging_2012}
\bibinfo{author}{Fung, J.}, \bibinfo{author}{Perry, R.~W.},
  \bibinfo{author}{Dimiduk, T.~G.} \& \bibinfo{author}{Manoharan, V.~N.}
\newblock \bibinfo{title}{Imaging multiple colloidal particles by fitting
  electromagnetic scattering solutions to digital holograms}.
\newblock \emph{\bibinfo{journal}{Journal of Quantitative Spectroscopy and
  Radiative Transfer}} \textbf{\bibinfo{volume}{113}},
  \bibinfo{pages}{2482--2489} (\bibinfo{year}{2012}).

\bibitem{hell_1993_aberrations}
\bibinfo{author}{Hell, S.}, \bibinfo{author}{Reiner, G.},
  \bibinfo{author}{Cremer, C.} \& \bibinfo{author}{Stelzer, E.~H.}
\newblock \bibinfo{title}{Aberrations in confocal fluorescence microscopy
  induced by mismatches in refractive index}.
\newblock \emph{\bibinfo{journal}{Journal of microscopy}}
  \textbf{\bibinfo{volume}{169}}, \bibinfo{pages}{391--405}
  (\bibinfo{year}{1993}).

\bibitem{visser_1992_refractive}
\bibinfo{author}{Visser, T.}, \bibinfo{author}{Oud, J.} \&
  \bibinfo{author}{Brakenhoff, G.}
\newblock \bibinfo{title}{Refractive index and axial distance measurements in
  3-d microscopy}.
\newblock \emph{\bibinfo{journal}{Optik}} \textbf{\bibinfo{volume}{90}},
  \bibinfo{pages}{17--19} (\bibinfo{year}{1992}).

\bibitem{fischer_stiffness-controlled_2012}
\bibinfo{author}{Fischer, R.~S.}, \bibinfo{author}{Myers, K.~A.},
  \bibinfo{author}{Gardel, M.~L.} \& \bibinfo{author}{Waterman, C.~M.}
\newblock \bibinfo{title}{Stiffness-controlled three-dimensional extracellular
  matrices for high-resolution imaging of cell behavior}.
\newblock \emph{\bibinfo{journal}{Nature Protocols}}
  \textbf{\bibinfo{volume}{7}}, \bibinfo{pages}{2056--2066}
  (\bibinfo{year}{2012}).

\bibitem{Halldorsson2015}
\bibinfo{author}{Halldorsson, S.}, \bibinfo{author}{Lucumi, E.},
  \bibinfo{author}{R., G.-S.} \& \bibinfo{author}{Fleming, R.}
\newblock \bibinfo{title}{Advantages and challenges of microfluidic cell
  culture in polydimethylsiloxane devices.}
\newblock \emph{\bibinfo{journal}{Biosensors and Bioelectronics}}
  \textbf{\bibinfo{volume}{63}}, \bibinfo{pages}{218--231}
  (\bibinfo{year}{2015}).

\end{thebibliography}

\clearpage

\clearpage

\clearpage

\end{document}